\documentclass{ws-p8-50x6-00}
\begin{document}
\title{Gravitational Experiment Below 1 Millimeter and
  Comment on Shielded Casimir Backgrounds for Experiments in the Micron Regime}
\author{Joshua C. Long, Allison B. Churnside, and John C. Price}
\address{University of Colorado, Boulder CO 80309 USA}

\maketitle
\abstracts{We present the status of an experimental test for
  gravitational strength forces below 1 mm.  Our experiment uses small
  1 kilohertz oscillators as test masses, with a stiff counducting
  shield between them to suppress backgrounds.  At the present sensitivity
  of approximately $10^3$ times gravitational strength, we see no
  evidence for new forces with interaction ranges between 
  75 $\mu$m and 1 mm.  While the Casimir background is not expected to
  be significant at this range, an extension of the shielding 
  technique we employ may be useful for reducing this background
  in experiments below a few microns.  We describe a possible implementation.} 

\section{Introduction}
Experimental searches for new macroscopic forces have only marginally
explored the distance range under 1 mm and there is little 
knowledge of gravity itself in this range.  The sub-millimeter region
is of profound and rapidly increasing experimental interest, given a 
number of recent predictions of new forces from a rich variety of 
modern theories of fundamental interactions.  

The existing experimental limits on new forces at short distances are
defined by classical gravity measurements and Casimir force
measurements,\cite{rev_ex} as shown in Fig.~\ref{fig:limit}.  
From the figure, in which the strength $\alpha$ of a hypothetical 
new force relative to gravity is plotted versus the Yukawa range 
$\lambda$, the published experimental limits allow for forces in nature 
several million times stronger than gravity over ranges as great as 
100 $\mu$m.

Also shown in Fig.~\ref{fig:limit} are theoretical predictions of new
effects in this regime.\cite{rev_th}  Most notable is the line indicating
Yukawa corrections to the inverse square law which arise from compact 
extra dimensions.\cite{dimensions}  Corrections are also predicted 
from massive scalars in string theories, such as Moduli and 
Dilatons.\cite{scalars}  The other predictions shown are motivated by the
Cosmological Constant Problem\cite{cosmological} and the Strong CP Problem
of QCD.\cite{pseudo} 

Since the summer of 1997, we have been constructing an experiment to 
explore the sub-centimeter regime.  The approach uses a high-frequency
technique, departing from the torsion balance method which was
used in all measurements defining the current limits. 

\section{Status of the Experiment}
\subsection{Design Overview}
\noindent  If all dimensions of experimental test masses of a
gravitational experiment are scaled by the same factor $D$, Newtonian
attraction varies as $D^{4}$.  At small separations, signals from
mass--coupled forces become very weak.  At the same time, background
effects such as surface forces increase rapidly.

\begin{figure}[htbp]
\begin{center}
\epsfxsize=25pc \epsfbox{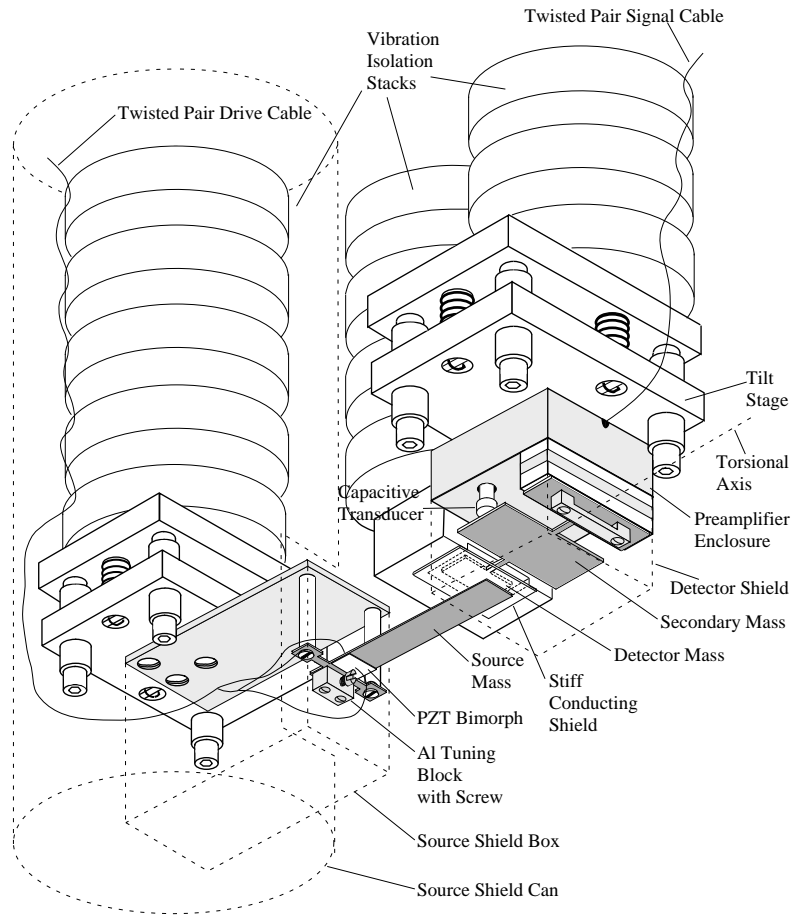}
\caption{Central components of the apparatus. \label{fig:expt}}
\end{center}
\end{figure}

Our experimental approach is shown in Fig.~\ref{fig:expt} (see
Ref.~\ref{ref:rev_ex} and the last paper of Ref.~\ref{ref:dimensions}
for more detail).  Planar
test mass geometry is chosen to concentrate as much mass as possible
at the scale of interest.  The test masses consist of 200 $\mu$m thick 
tungsten oscillators.  The detector is driven by the source
mass on resonance near 1 kHz.  At this frequency it is possible 
to construct simple, passive vibration isolation stacks sufficient 
to suppress the acoustic coupling of the test masses though the
apparatus.  Detector oscillations are read out with a capacitive 
transducer and lock-in amplifier.  The entire apparatus is enclosed 
in a vacuum chamber and operated at $10^{-7}$ torr to reduce the 
acoustic coupling between the test masses though the residual gas 
in between them.

The principal backgrounds in addition to the acoustic backgrounds
arise from electrostatic and magnetic forces.  Electrostatic forces,
as well as acoustic coupling through the residual gas,
are suppressed with a stiff conducting shield, consisting of a
75 $\mu$m thick quartz window plated with gold, suspended between the 
test masses.  Magnetic backgrounds have so far been avoided with the
exclusive use of non-magnetic materials for the construction of the
apparatus.  If the need arises, in-situ imaging of magnetic
contaminants on the test masses is possible. 

If all of these backgrounds can be sufficiently reduced, the limiting
background of this experiment will be thermal noise due to dissipation
in the detector mass.  The experiment is therefore being carried out
in two phases.  A preliminary, room--temperature version of
the experiment is now in operation and nearing the practical limit of its
sensitivity.  The final version of the experiment will be cooled to 
liquid Helium temperatures for greater sensitivity.

\subsection{Current Sensitivity}
A signal from a new effect can be modeled as the force on the detector
mass due to a Yukawa interaction with the source mass. This force is
given by:
\begin{equation}
F_{Y}(t) = 2 \pi \alpha G \rho_{s} \rho_{d} A \lambda^{2} 
\exp{(-d(t)/ \lambda)}[1 -\exp{(-t_{s} / \lambda)} ] 
[1 - \exp{(-t_{d} / \lambda)}],
\label{eq:FYp}
\end{equation}
where $d(t)$ is the separation between the test masses, $\rho_{s}$
and $\rho_{d}$ are the source and detector mass densities, and $t_{s}$
and $t_{d}$ are the thicknesses.  This would be
an exact expression if either plate had area $A$ and the other had 
infinite area, but for the real geometry there are small edge 
corrections, which we do not consider.  For values of these parameters
typical for our experiment, this force is approximately $2 \times
10^{-14}$ N, for $\alpha = 1$ and $\lambda$ = 100 $\mu$m. 

The rms thermal noise force is found from the mechanical Nyquist theorem to be:
\begin{equation}
F_{T} = \sqrt{\frac{4kT}{\tau} \left( \frac{m\omega}{Q} 
\right)},
\label{eq:FT}
\end{equation}
where $m$ is the mass of the small rectangular section of the detector
oscillator, $\omega$ is the resonant frequency, $Q$ is the detector 
quality factor, $T$ is the temperature and $\tau$ is the 
measurement integration time.  For typical $Q$ values of our tungsten
detector mass ($2.5 \times 10^4$), a temperature of 300 K, and an
integration time of 1000 s, this force (and hence the current
sensitivity of our experiment) is about $4 \times 10^{-14}$ N. 

\begin{figure}[htbp]
\begin{center}
\epsfxsize=25pc \epsfbox{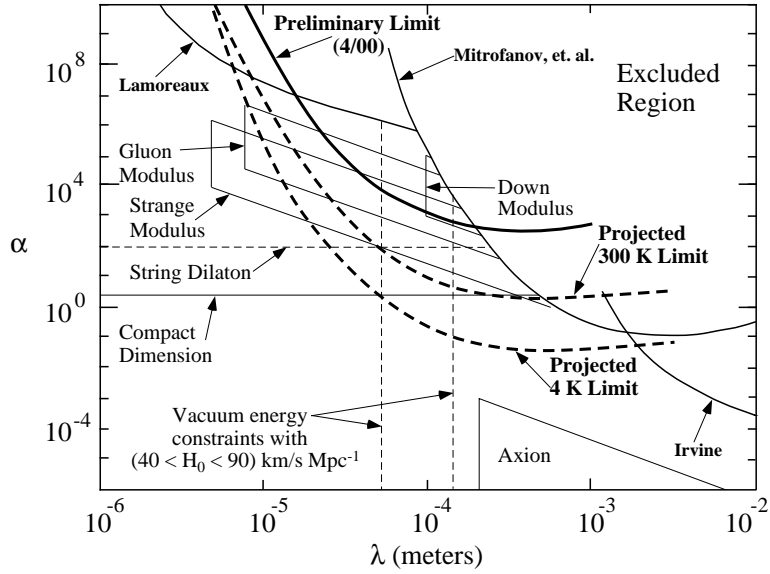}
\caption{Parameter space for Yukawa-type forces in which the strength
  relative to gravity ($\alpha$) is plotted versus the range
  ($\lambda$).  Limit curves from published experiments are shown along
  with theoretical predictions of new phenomena in this regime.  The
  bold lines indicate the current (solid) and projected (dashed) sensitivities
  of our experiment. \label{fig:limit}}
\end{center}
\end{figure}

In April 2000, the room temperature experiment became fully
operational for the first time.  No signal was observed above the
detector thermal noise over the entire integration time of 1800 s.
Setting the ratio of Eqs.~\ref{eq:FYp} and~\ref{eq:FT} to unity and
solving for $\alpha$, we infer the limits on the Yukawa parameters as
shown in Fig.~\ref{fig:limit}.

\subsection{Recent Improvements}
For the April run, thermal drifts of the detector resonance limited
the integration time to 1800 s.  Furthermore, a poor match between source
and detector resonant frequencies limited the source mass amplitude to
roughly 0.5 $\mu$m.  Since that run we have installed temperature
control, allowing for much longer integration times.  We have
tuned the source mass such that amplitudes greater than 50 $\mu$m on
resonance are now possible.  Also, we have fabricated a new shield
from 50 $\mu$m thick sapphire, making smaller separations
possible.  With these improvements, we expect the room temperature experiment
to attain the sensitivity
represented by the upper dashed curve in Fig.~\ref{fig:limit}.

The final cryogenic version of the experiment is under development.  
Assuming the backgrounds can be controlled, we expect
this experiment to reach gravitational strength at 50 $\mu$m, as 
represented by the lower dashed curve in Fig.~\ref{fig:limit}.

\section{Shielded Casimir Background in the Micron Regime}
Session PT6 at MG9 featured discussions of experimental searches for 
new forces in the sub-micron regime, where the Casimir background is 
expected to dominate.  A series of Casimir force measurements has 
been carried out in this regime using an AFM,\cite{afm} and 
limits on new forces down to 10 nm have been derived from these 
experiments.\cite{casimir_limits}  So far, the limits are many 
orders of magnitude stronger than gravity, 
motivating the design of dedicated searches for new forces.  One
proposal in this regime is the ``iso-electronic'' 
technique of Fischbach and Krause.\cite{isoelectronic}  

The Casimir background is not expected to be significant in our 
existing experiment.  However, an extension of the shielding technique
we employ may be useful for AFM--type searches in experiments near 1 $\mu$m.

\begin{figure}[htbp]
\begin{center}
\epsfxsize=25pc \epsfbox{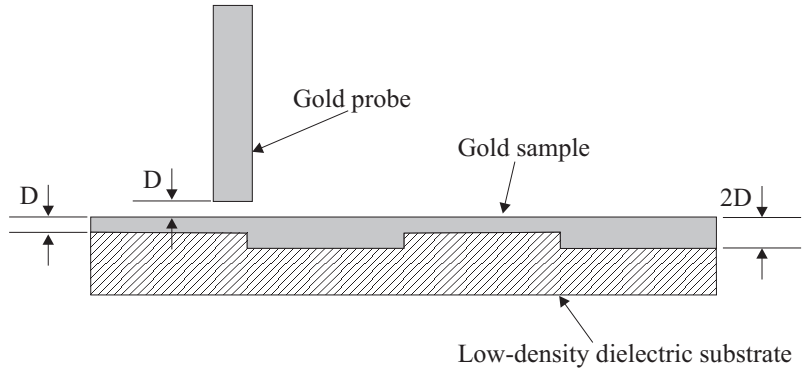}
\caption{Idealized experimental geometry.  Horizontal dimensions
  of all features are taken to be much larger than $D$.  
  \label{fig:afm}}
\end{center}
\end{figure}

Our idea is illustrated in Fig.~\ref{fig:afm}.  A probe at a 
distance $D$ above a flat gold sample is scanned over regions of 
alternating depth $D$ and 2$D$. The horizontal dimensions of all probe
and sample features are taken to be much larger than $D$.
The gold sample is backed by a low--density dielectric substrate
which for the purposes of this study we take to be vacuum.

Due to the finite penetration depth of the sample metal, above a
certain scale $D$ we expect the Casimir force between the probe 
and either thickness region of the sample to be essentially
equal.  While mass--coupled forces between probe and sample may be
much more feeble, for them we do not expect a similar equalization across the
thickness boundaries.

We have calculated the Casimir forces for the geometry in Fig~\ref{fig:afm}.
We use the method developed by A. Lambrecht and S. 
Reynaud,\cite{mirrors} ignoring the effects of surface roughness and
finite temperature.  We employ a Drude model for the gold probe 
and sample, using a plasma frequency of $\omega_{P} = 1.4 
\times 10^{16}$ rad/s and relaxation parameter of $\gamma = 5.3 
\times 10^{13}$ rad/s in Eq.~16 of Ref.~\ref{ref:mirrors}.  

Setting the plate separation $L$ in Eq.~9 of Ref.~\ref{ref:mirrors}
equal to $D$, we then use Eqs.~12, 9, and 4 of that reference to 
calculate the difference in the Casimir force between the probe 
and sample regions of thickness $D$ and 2$D$: 
$\Delta F_{C} = F_{C}^{2D}-F_{C}^{D}$.  We assume the probe to have 
infinite thickness.  As a preliminary check, we set the sample
thickness equal to infinity and note that our results for the Casimir
reduction factor $\eta_{F}$ agree with those of Fig.~6 of 
Ref.~\ref{ref:mirrors} to within 5\%.

\begin{figure}[htbp]
\begin{center}
\epsfxsize=25pc \epsfbox{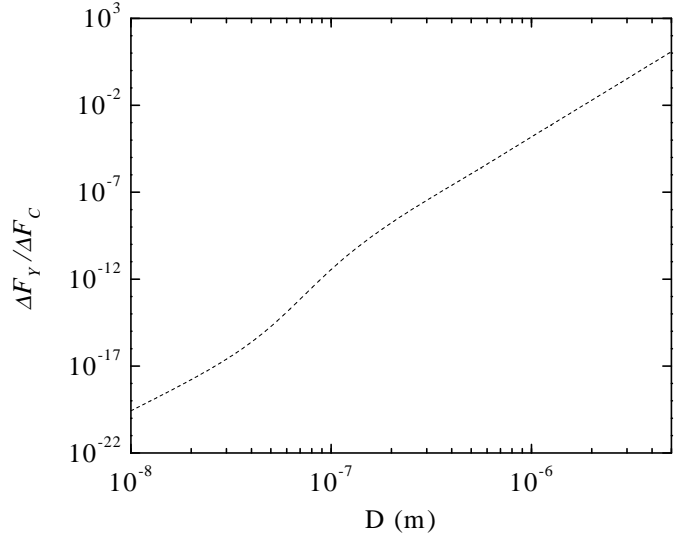}
\caption{Scale dependence of variation of Yukawa force with $\alpha =
  1$, $\lambda = D$, relative to variation of Casimir Force as probe 
  is scanned from over sample region of thickness $D$ to region of 
  thickness $2D$. \label{fig:fyfc}}
\end{center}
\end{figure}

We then compute the difference in the Yukawa force for the same 
geometry, using $\alpha = 1$ and $\lambda = D$: 
$\Delta F_{Y} = F_{Y}^{2D}-F_{Y}^{D}$.  
The ratio $\Delta F_{Y}/\Delta F_{C}$ is shown in Fig.~\ref{fig:fyfc}
as a function of $D$ for $10^{-8}$ m $< D < 5 \times 10^{-6}$ m.  
We note that these results change by less than 1\% as we vary the 
limits of integration [$10^{8}$--$10^{20}$ rad/s] by an order of magnitude.

The steep rise in the curve occurs when the scale $D$ approaches
the plasma wavelength we have used for gold (about $1.4 \times
10^{-7}$ m), at which point the shielding increases rapidly.
For the idealized geometry and conditions we have assumed, Fig.~\ref{fig:fyfc}
implies that a gravitational--strength Yukawa force becomes
distinguishable from the Casimir background as the probe
is scanned across the thickness boundary at a scale $D$ of about 3 $\mu$m.  

\section*{Acknowledgments}
We wish to thank Hilton Chan for his extensive work on
this project.  J. L. thanks A. Lambrecht and S. Reynaud for
useful discussions.

\end{document}